\documentclass[aps,pre,twocolumn,showpacs,superscriptaddress]{revtex4}
\usepackage{graphicx,color,epsfig}
\usepackage{amssymb,amsmath,amsthm}
\begin{document}
\title{Equivalent synchronization of chaos in driven and in autonomous systems} 
\author{M. G. Cosenza}
\affiliation{Centro de F\'isica Fundamental, Universidad de Los Andes, M\'erida, Venezuela}
\author{O. Alvarez-Llamoza} 
\affiliation{Departamento de F\'isica, FACYT, Universidad de Carabobo, Valencia, Venezuela}
\author{G. Paredes}
\affiliation{Laboratorio de F\'isica Aplicada y Computacional, Universidad Nacional Experimental del T\'achira, San Crist\'obal, Venezuela}

\begin{abstract}
It is shown that the synchronization behavior of a system of chaotic maps subject to either an external forcing or a coupling function of their internal variables can be inferred from the behavior of a single element in the system, which can be seen as a single drive-response map. 
From the conditions for stable synchronization in this single driven-map model with minimal ingredients, we find minimal conditions for the emergence of complete and generalized chaos synchronization in both driven and autonomous associated systems. Our results show that the presence of a common drive or a coupling function for all times is not indispensable for reaching synchronization in a system of chaotic oscillators, nor is the simultaneous sharing of
a field, either external or endogenous, 
by all the elements. In the case of an autonomous system, the coupling function does not need to depend on all the internal variables for achieving synchronization and its functional form is not crucial for generalized synchronization.  What becomes essential for reaching synchronization in an extended system is
the sharing of some minimal information by its elements, on the average, over long times, independently of the nature (external or internal) of its source. 
\end{abstract}
\pacs{05.45.-a, 05.45.Xt, 05.45.Ra}
\date{\today}
\maketitle

\section{Introduction}
Chaos synchronization is a fundamental phenomenon in dynamical systems (Pecora \& Carroll 1990, Pikovsky {\it et al.} 2002). 
Its study has provided insights into many natural processes and motivation for practical
applications such as secure communications and control of dynamical systems (Boccaletti {\it et al.} 2002, Uchida {\it et al.} 2005, Argyris
{\it et al.} 2005).
Chaos synchronization is commonly found in unidirectionally coupled systems, where a distinction can be made between 
a drive or forcing subsystem and another driven or response subsystem that possesses chaotic dynamics (Pikovsky {\it et al.} 2002).
Complete synchronization occurs when the state variables 
of the drive and the response subsystems converge to a single orbit in phase space.
More generally, generalized synchronization of chaos arises when a functional relation different from the identity is established between the drive and the response subsystems (Rulkov et al 1995, Abarbanel {\it et al.} 1996, Kapitaniak {\it et al.} 1996, Hunt {\it et al.} 1997, Parlitz \&  Kokarev 1999, Zhou \& Roy 2007).

The auxiliary system approach (Abarbanel {\it et al.} 1996) shows that when a response and a replica subsystems are driven by the same
signal, then the orbits in the phase spaces of the
response and replica subsystems become identical and they
can evolve on identical attractors, if their initial conditions lie on the same basin of attraction of the driven-response system. By extension, an ensemble of identical chaotic oscillators can also be synchronized by a common drive. In this case, generalized synchronization implies that the elements in the ensemble become synchronized into a single trajectory that evolves differently from that of the drive. The specific functional form of the drive is not determinantal; the basic mechanism that leads to synchronization is the sharing of the same input by the oscillators for all times. 

On the other hand, there has been interest in the investigation of chaotic synchronization and other forms of collective behaviors emerging in autonomous systems consisting of networks of mutually interacting nonlinear units without the presence of external influences (Newman {\it et al.} 2006, Manrubia {\it et al.} 2004). 
Synchronization in globally coupled oscillators is relevant in many chemical and biological systems and have been experimentally investigated (Wang {\it et al.} 2000, S. De Monte {\it et al.} (2007), Taylor {\it et al.} 2009). 
In autonomous systems where the units are coupled through their mean field, the units and the mean field can synchronize to a common chaotic trajectory (Kaneko 1989). This process can be regarded as the analogous of complete synchronization in driven systems of elements subject to a common drive. Recently, the concept of generalized synchronization of chaos occurring in driven systems has been applied to the context of autonomous systems (Alvarez-Llamoza \& Cosenza 2008). This means that, under some circumstances, the chaotic state variables in an autonomous dynamical system can be synchronized to each other but not to a shared coupling function containing information from those variables. 

The occurrence of both forms of chaos synchronization in driven and in autonomous systems suggests that 
the nature of the common influence acting on the elements in a system is irrelevant; it could consist of an external forcing, or an autonomous interaction function (Alvarez-Llamoza \& Cosenza 2008). At the local level, each element in the system is subject to a field, either external or endogenous, that eventually induces some form of synchronization between that field and the element. 
Thus, the local dynamics can be seen as a single element subject to a drive evolving as the common field in each case. 

In this paper we present a simple and unified scheme for the study of chaos synchronization in classes of driven and in autonomous dynamical systems. 
Our method is based on the reported analogy between a single driven map and a system of globally coupled maps (Parravano \& Cosenza 1998). This analogy provided an explanation of dynamical clustering  in globally coupled systems (Parravano \& Cosenza 1998, Manrubia {\it et al.} 2004) and of stability of steady states in systems with delayed interactions (Masoller \&  Marti 2005).
We show that the synchronization behavior of a system of chaotic maps subject to either an external forcing or a coupling function of their internal variables can be inferred from the behavior of a single driven map in the system.  When this single drive-response system reaches either complete or generalized synchronization, the local analogy implies that an ensemble of identical maps subject to a common (external or internal) field that behaves as this drive, should also synchronize in a similar fashion. We show that the conditions for stable synchronization in a single driven map also describe the regions where synchronized states occur in both driven and autonomous extended systems. 

By studying a single driven-map model with minimal ingredients, we find minimal conditions for the emergence of chaos synchronization in both driven and autonomous associated systems. 
Our results show that the presence of a common drive  or a coupling function for all times is not indispensable for reaching synchronization in an extended system of chaotic oscillators, nor is the simultaneous sharing of the drive or the coupling function by all the elements in the system. In the case of an autonomous system, the coupling function does not need to depend on all the internal variables for achieving synchronization and, in particular, its functional form is not crucial for generalized synchronization. What seems essential for reaching synchronization in a system is
the sharing of some minimal information by the elements, on the average, over long times, independently of its source.

In Sec.~II, the single driven map model is presented and the stability conditions for complete and generalized chaotic synchronization are analyzed. The synchronization behavior of a system of homogeneously driven maps is explored in Sec.~III
and that of a system of maps subject to an asynchronous drive is shown in Sec.~IV. An autonomous system of homogeneously coupled maps is considered in Sec.~V. The behavior of an autonomous system of heterogeneously coupled maps is shown in
Sec.~VI. Conclusions are presented in Sec.~VII.

\section{Single driven map}
Our work is motivated by the practical aspect of searching for minimal requirements for the emergence of chaos synchronization in dynamical systems.
As a minimal model, 
we consider a single map driven with a probability $p$,
\begin{equation}
\begin{array}{l}
x_{t+1} = \begin{cases}
w(x_t,y_t), & \text{with probability $p$}\\
f(x_t),  & \text{with probability $(1-p)$},
          \end{cases} \\
y_{t+1} =  g(y_t),
\end{array}
\label{2dmap}
\end{equation}
where $f(x_t)$ and $g(y_t)$ describe the dynamics of the driven and the drive variables, $x_t$ and $y_t$, respectively, at discrete time $t$ and the coupling relation between them is chosen 
to be of the diffusive form 
\begin{equation}
  w(x_t,y_t)=(1-\epsilon)f(x_t)+ \epsilon g(y_t) \, ,
\end{equation}
where $\epsilon$ is the coupling strength.
Since we are focused on chaotic synchronization, 
for the driven dynamics we shall choose $f(x_t)=b+ \ln | x_t |$,
where $b$ is a real parameter and $x_t \in (-\infty, \infty)$. This logarithmic
map exhibits robust chaos, with no periodic windows and no separated chaotic
bands, on the interval $b \in [-1,1]$ (Kawabe \& Kondo 1991). Robustness is an important property in practical applications requiring reliable operation under chaos in the sense that the chaotic behavior cannot be modified by small perturbations of the system parameters.

The linear stability condition for synchronization is determined by the Lyapunov exponents of the
two-dimensional system Eq.~(\ref{2dmap}). These are defined as $\Lambda_x = \lim_{T\rightarrow \infty} \ln L_x$ and $\Lambda_y = \lim_{T\rightarrow \infty} \ln L_y$, where $L_x$ and $L_y$ are the magnitude of the eigenvalues of $[\prod^{T-1}_{t=0} \mathbf{J}(x_t,y_t)] ^{1/T}$, and $\mathbf{J}(x_t,y_t)$ is the Jacobian matrix for the system  Eq.(\ref{2dmap}), calculated along an orbit. A given orbit $\{x_t,y_t\}$ from $t=0$ to $t=T-1$ can be separated in two subsets,
according to the source of the $x_t$ variable, either coupled or uncoupled, that we respectively denote as 
$A=\{ \{x_t,y_t\}:  x_t=w(x_{t-1},y_{t-1}) \}$ possessing $pT$ elements, and $B=\{ \{x_t,y_t\}: x_t=f(x_{t-1}) \}$ having $(1-p)T$ elements. We get
\begin{equation}
\left( \prod^{T-1}_{t=0} \mathbf{J}\right)^{1/T} = \left( \begin{array}{cc}
\displaystyle{ \prod_{t: \, x_t \in A} w_x \prod_{t: \, x_t \in B} f'(x_t)} & K\\
0 & \displaystyle{\prod^{T-1}_{t=0} g'(y_t)}
\end{array} \right)^{1/T},
\label{jacobians2}
\end{equation}
where $w_x=\frac{\partial w}{\partial x}=(1-\epsilon)f'(x)$, and $K$ is a polynomial whose terms contain products of $w_x$, $\epsilon$, and $g'(y_t)$ to be evaluated along time.
Then $L_x=[ \prod_{x_t \in A} w_x \prod_{x_t \in B} f'(x_t)]^{1/T}$
and $L_y=[ \prod^{T-1}_{t=0} g'(y_t) ]^{1/T}$. Thus we get 
\begin{equation}
\Lambda_x= p \ln|1-\epsilon| + \lim_{T \rightarrow \infty} \dfrac{1}{T}  \left[ \ln \prod_{x_t \in A} \vert f'(x_t)\vert + \ln  \prod_{x_t \in B} \vert f'(x_t)\vert \right]              
\end{equation}
\begin{equation}
 \Lambda_y =\lim_{T \rightarrow \infty} \dfrac{1}{T} \sum^{T-1}_{t=0} \ln | g'(y_t) | = \lambda_g \, ,
\end{equation}
where $\lambda_g$ is the Lyapunov exponent of the map $g(y_t)$. We shall assume chaotic drives $g(y_t)$; thus $\Lambda_y >0$. Synchronization takes place when the Lyapunov exponent corresponding to the driven map is negative (Rulkov {\it et al.} 1995); i.e., $\Lambda_x<0$. For a given set of parameter values, there is a definite value of the probability $p$ at which the exponent $\Lambda_x$ may change its sign, from positive to negative, signaling the onset of synchronization  and the appearance of a contracting direction in the dynamics of the single driven map system Eq.~(\ref{2dmap}). Thus, the drive just needs to act a fraction $p$ of the time in order to achieve synchronization in the system Eq.~(\ref{2dmap}).

When $g=f$, the condition $\Lambda_x<0$ implies complete synchronization, where $x_t=y_t$. In this case we get 
\begin{equation}
{\begin{array}{l}
\Lambda_y= \lambda_f \, , \\
\Lambda_x= p \ln |1-\epsilon| + \lambda_f \, ,
 \end{array}}
\label{lyapexp}
\end{equation}
where $\lambda_f$ is the Lyapunov exponent of the map $f$.

\begin{figure}[h]
\centerline{
\includegraphics[scale=0.54]{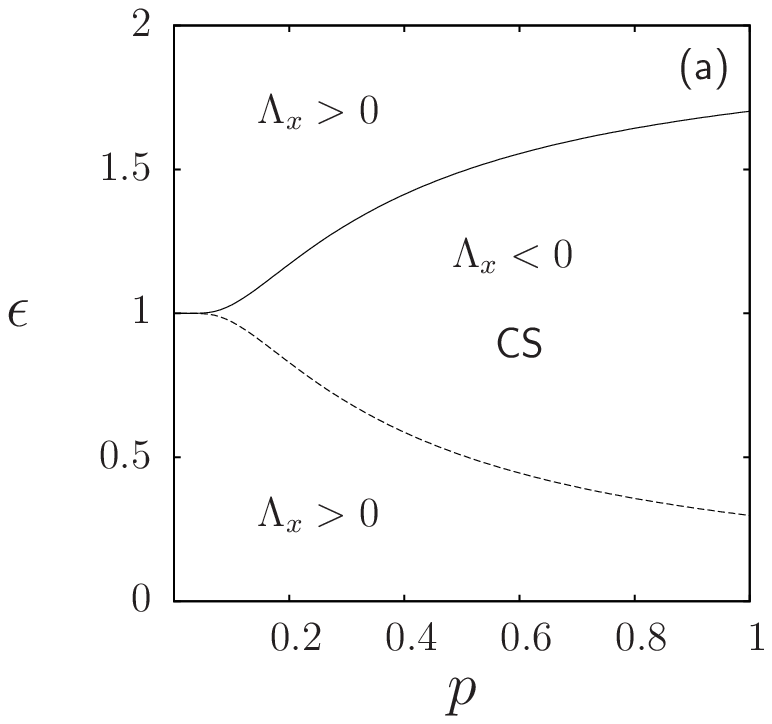}
\hspace{1mm}
\includegraphics[scale=0.54]{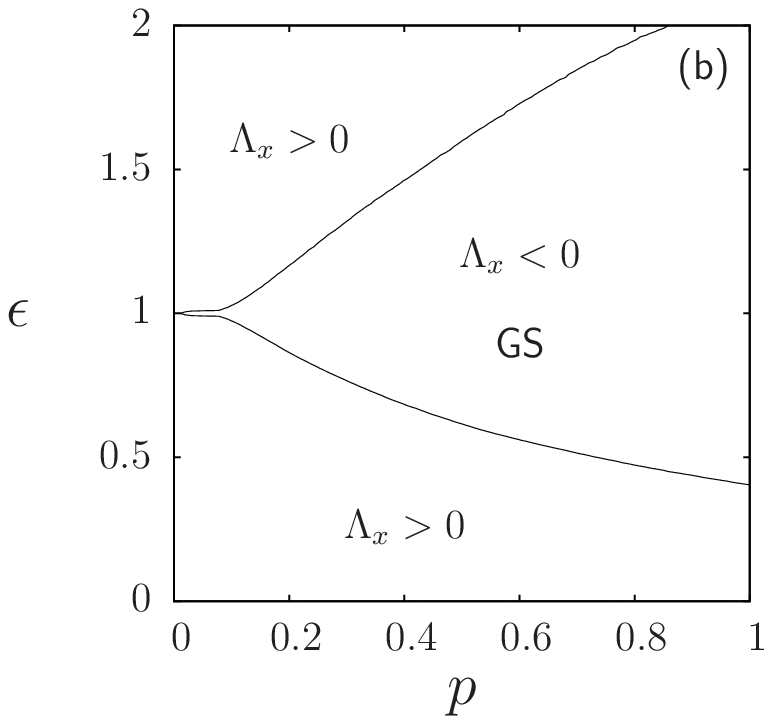}}
\caption{Boundaries $\Lambda_x = 0$ on the plane $(p,\epsilon)$ for synchronization
in the  single driven map Eq.~(\ref{2dmap}) with $f=-0.7+ \ln | x_t |$.
(a) Complete synchronization (CS) with $g=f=-0.7+ \ln | y_t |$.
(b) Generalized synchronization (GS) with $g(y_t)=0.5+ \ln | y_t | \neq f$.}
\label{fig1}
\end{figure}

Figure~\ref{fig1}(a) shows the stability boundary, given by $\Lambda_x=0$, for the completely synchronized states of the system  Eq.~(\ref{2dmap}) on the space of parameters $(p,\epsilon)$ for a drive $g=f$.
On the other hand, if $g\neq f$, the condition $\Lambda_x<0$ corresponds to generalized synchronization, characterized by $x_t \neq y_t$. Figure~\ref{fig1}(b) shows the stability boundary, given by $\Lambda_x=0$, for the generalized synchronized states of the system Eq.~(\ref{2dmap}) on the space of parameters $(p,\epsilon)$ for a drive $g \neq f$.

\section{Homogeneously driven systems}
The auxiliary system approach (Abarbanel {\it et al.} 1996) implies that a driven map can synchronize on identical orbits with another, identically driven map. By extension, let us consider a system of $N$ intermittently and homogeneously driven maps, defined as
\begin{equation}
\label{int_drive}
\forall i, \;  x^i_{t+1} = \left\lbrace 
\begin{array}{ll}
(1-\epsilon)f(x^i_t)+ \epsilon g(y_t), &    \text{with prob. $p$},\\
f(x_t^i),  &   \text{with prob. $(1-p)$} 
\end{array}
\right. 
\end{equation}
where $x^i_t$ ($i=1,2,\ldots,N$) gives the state of the $i$th
map at discrete time $t$, $f$ describes the local dynamics, and $\epsilon$ is the strength of the coupling to the drive $g(y_t)$.

Since each map in the system Eq.~(\ref{int_drive}) is subject to the same external influence (or lack of it) at any time, the properties of this system can be characterized from the behavior of the individual local dynamics. 
Thus, the occurrence of stable synchronization in the single driven map Eq.~(\ref{2dmap}) should lead to synchronization in the extended system of maps Eq.~(\ref{int_drive}), even when the drive acts intermittently in both cases.
A completely synchronized state in the system Eq.~(\ref{int_drive}) is given by $x_t^i=x_t=y_t$, $\forall i$, and it can occur when $g=f$. On the other hand, if $g\neq f$, generalized synchronization, characterized by the condition $x_t^i=x_t \neq y_t$,
$\forall i$, may also arise in this system for $p \leq 1$.

Synchronization in an extended system can be characterized by the asymptotic time-average $\langle\sigma\rangle$ (after discarding a number of transients) of the instantaneous standard deviations
$\sigma_t$ of the distribution of state variables $x^i_t$, defined as
\begin{equation}
\sigma_t=\left[ \frac{1}{N} \sum_{i=1}^N \left( x^i_t - S_t \right)^2 \right]^{1/2},
\end{equation}
where $S_t$ is 
the instantaneous mean field of the system,
\begin{equation}
\label{mean}
 S_t=\frac{1}{N} \sum^N_{i=1} x^i_t \, .
\end{equation}
Stable synchronization corresponds to $\langle \sigma \rangle=0$. Here we use the numerical criterion 
$\langle \sigma \rangle < 10^{-7}$.

\begin{figure}[h]
\begin{center}
\includegraphics[scale=0.72]{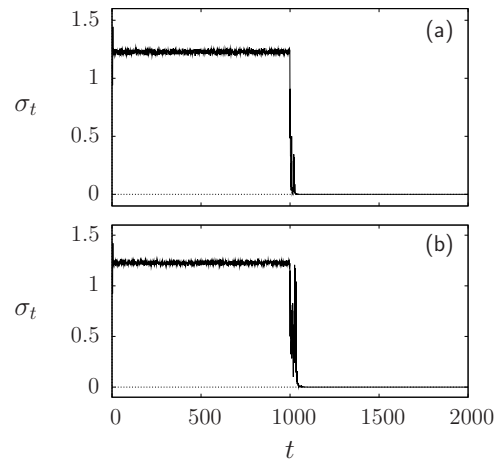}
\caption{\label{fig2} $\sigma_t$ vs. $t$ for the systems of driven maps
Eq.~(\ref{int_drive}) with $f(x_t^i)=-0.7+ \ln | x_t^i |$, for different forms of the drive $g(y)$. (a) $g=-0.7+ \ln|y_t|=f$; complete synchronization. (b) $g(y_t)=0.5+\ln | x_t | \neq f$; generalized synchronization. 
In both cases, $N=10^4$, $\epsilon=0.6$, $p=0.6$, initial conditions were randomly and uniformly distributed such that $x_i \in [-10,3]$ and the drive $g(y_t)$ is applied starting at $t=1000$.}
\end{center}
\end{figure}

Figure~\ref{fig2} shows $\sigma_t$  as a function of time for the intermittently driven system Eq.~(\ref{int_drive}) subject to  chaotic drives $g=f$ and $g\neq f$, corresponding to complete and generalized synchronization, respectively.

For a given value of the coupling strength $\epsilon$, there is a threshold value of the probability $p$ required to reach
either type of synchronization. In each case, the condition $\Lambda_x < 0$ for synchronization in the single driven map Eq.~(\ref{2dmap}) implies stable complete or generalized synchronization in the intermittently driven system  Eq.~(\ref{int_drive}). For the same values of parameters as in the single driven map, the region for the complete synchronized state on the space of parameters $(p,\epsilon)$  of the system Eq.~(\ref{int_drive}) with a drive $g=f$ is the same as in Figure~\ref{fig1}(a). Similarly, Figure~\ref{fig1}(b) describes the region for generalized synchronization corresponding to the system Eq.~(\ref{int_drive}) subject to a drive $g \neq f$.

\section{Heterogeneously driven systems}
The equivalence between 
a single driven map and a system of driven similar maps also suggests that, under some circumstances, the collective behavior of an extended system of interacting elements can be inferred by considering the dynamics of a single element at the local level. As an application of this idea, we consider a system of heterogeneously driven maps defined as
\begin{equation}
\label{part_drive}
x^i_{t+1} = \left\lbrace 
\begin{array}{ll}
(1-\epsilon)f(x^i_t)+ \epsilon g(y_t), & \text{with probability $p$},\\
f(x_t^i), & \text{with probability $(1-p)$.} 
\end{array}
\right. 
\end{equation}
The parameter $p$ is the probability of interaction of a map with the drive $g$ at a time $t$. 
The driven elements are randomly chosen with a probability $p$, so that not all the maps in the system receive the same external influence at all times. Thus, the average fraction of driven elements in the system Eq.~(\ref{part_drive}) at 
any given time is $p$. In comparison, the forcing of the elements in the homogeneously driven system Eq.~(\ref{int_drive}) is simultaneous and uniform; each map receives the same influence from the drive $g$ at any $t$ with probability $p$.

\begin{figure}[h]
\centerline{
\includegraphics[scale=0.75]{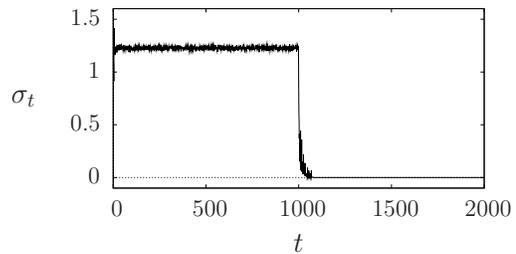}}
\caption{$\sigma_t$ vs. $t$ for the systems of driven maps
Eq.~(\ref{part_drive}) with $f(x_t^i)=-0.7+ \ln | x_t^i |$, for $g=-0.7+ \ln|y_t|$; complete synchronization. Parameters are $N=10^4$, $\epsilon=0.6$, $p=0.6$; initial conditions were randomly and uniformly distributed such that $x_i \in [-10,3]$ and the drive $g(y_t)$ is applied starting at $t=1000$.}
\label{fig3}
\end{figure}

When the heterogeneously driven system Eq.~(\ref{part_drive}) gets synchronized, we have $x_t^i=x_t$. However, the  synchronized solution exists only if $g=f$. Therefore, only complete synchronization $x_t^i=x_t=y_t$ can occur in this system. 
Figure~\ref{fig3} shows $\sigma_t$  as a function of time for the system Eq.~(\ref{part_drive}) subject to 
a drive $g=f$. 

At the local level, each map in the heterogeneously driven system Eq.~(\ref{part_drive}) is subject, over long times, to an external forcing $g$ with probability $p$. Thus, as in the homogeneously driven system  Eq.~(\ref{int_drive}), the synchronization behavior of system Eq.~(\ref{part_drive}) can also be studied from the
behavior of the single map driven with probability $p$, Eq.~(\ref{2dmap}). In particular, if the system of maps Eq.~(\ref{part_drive})
driven with $g=f$ reaches a complete synchronized state for some values of parameters, then for this same set of parameters the single driven map Eq.~(\ref{2dmap}) subject to the same drive should eventually exhibit a synchronized state similar to that of system Eq.~(\ref{part_drive}). 
Thus, the condition $\Lambda_x < 0$ for complete synchronization in the single driven map Eq.~(\ref{2dmap}), that implies
stable complete synchronization in the homogeneously driven system  Eq.~(\ref{int_drive}), 
also describes the stability of the complete synchronized state $x_t^i=x_t=y_t$, $\forall i$, in the heterogeneously driven system Eq.~(\ref{part_drive}). The stability boundary $\Lambda_x=0$ in Fig.~\ref{fig1}(a) for the driven map with $g=f$ coincides with both, the boundary that separates the region where complete synchronization occurs on the space of parameters $(p,\epsilon)$ for the homogeneously driven system Eq.~(\ref{int_drive}) and the boundary for complete synchronization in the  heterogeneously driven system Eq.~(\ref{part_drive}).

\section{Autonomous systems with homogeneous coupling}
The dynamics of the single driven map Eq.~(\ref{2dmap}) can also be compared
with the dynamics of an autonomous system of maps that share a global interaction with a probability $p$, 
\begin{equation}
\forall i, \; x^i_{t+1} = \left\lbrace 
\begin{array}{ll}
(1-\epsilon)f(x^i_t)+ \epsilon H( x^j_t:j \in Q_t), & \text{prob. $p$},\\
f(x_t^i), & \text{prob. $(1-p)$} 
\end{array}
\right. 
\label{auto_syst}
\end{equation}
where $\epsilon$ represents the magnitude of the coupling to the global interaction function $H$; and $Q_t$ is a subset having $q \leq N$ elements of the system that may be chosen at random at each time $t$. 
Each map receives the same information from the feedback coupling function $H$ at any $t$ with probability $p$.
When this autonomous system gets synchronized at some values of parameters, we have $x_t^i=x_t$. Thus for this same set of parameters, the single driven map subject to a forcing that satisfies $g(y_t)=H(x^j_t=x_t: j \in Q_t)$ for long times should exhibit a synchronized state similar to that of the autonomous system Eq.~(\ref{auto_syst}).
Thus, besides complete synchronization where $x_t^i=x_t=H$, 
other synchronized states, characterized by $x_t^i=x_t \neq H$ should also occur in the autonomous system, Eq.~(\ref{auto_syst}), for appropriate values of parameters. These states can be described as generalized synchronization in autonomous dynamical systems (Alvarez-Llamoza \& Cosenza 2008).

As an example of complete synchronization in the system Eq.~(\ref{auto_syst}), consider a partial mean field coupling function defined as
\begin{equation}
\label{partial}
 H( x^j_t:j \in Q_t)=\frac{1}{q} \sum^q_{j=1} f(x^j_t),
\end{equation}
where $q<N$ maps are randomly chosen at each time $t$. In this case the autonomous system Eq.~(\ref{auto_syst}) can be expressed in vector form as
\begin{equation}
 \mathbf{x}_{t+1} = 
   \begin{cases}
  \left( (1-\epsilon) \mathbf{I} + \dfrac{\epsilon}{q}\mathbf{G}_t \right)  \mathbf{f}(\mathbf{x}_t), \,
        \text{with probability $p$},\\
   \mathbf{I} ~ \mathbf{f}(\mathbf{x}_t), \quad \text{with probability $(1-p)$},
   \end{cases}
\label{aut_syst_vect}
\end{equation}
where the $N$-dimensional vectors $\mathbf{x}_t$ and $\mathbf{f}(\mathbf{x}_t)$
have components $[\mathbf{x}_t]_i=x_t^i$ and
$[\mathbf{f}(\mathbf{x}_t)]_i=f(x_t^i)$, respectively, 
$\mathbf{I}$ is the $N \times N$ identity matrix, and $\mathbf{G}_t$ is an $N\times N$ matrix 
that at each time $t$ possesses $q$ randomly chosen columns that have all their components equal to $1$ while the remaining $N-q$ columns have all their components equal to $0$.
The case $q=N$ and $p=1$ corresponds to the usual mean field global coupling (Kaneko 1989). The linear stability analysis (Waller \& Kapral 1884) of the complete synchronized state $f(x_t^i)=f(x_t)=H$ yields
\begin{equation}
\label{condition}
 \left| \left[ (1-\epsilon) + \frac{\epsilon}{q}\alpha_k \right]^p  e^{\lambda_f} \right| < 1 \, ,
\end{equation}
where $\alpha_k=\delta_{0k}q \,$ $(k=0,1\ldots,N-1)$ are the set of eigenvalues of the matrix $\mathbf{G}_t$ for any $t$, with the zero eigenvalue having $(N-1)$-fold degeneracy. The eigenvector corresponding to $k=0$ is homogeneous.
Thus only perturbations of $\mathbf{x}_t$ along the other 
eigenvectors may destroy the coherence.
Thus, condition Eq.~(\ref{condition}) with $k\neq 0$
becomes
\begin{equation}
p \ln |1-\epsilon |+ \lambda_f < 0,
\end{equation}
which is the same condition for stability of complete synchronized states in the driven map, Eq.~(\ref{lyapexp}), when $g=f$. Thus the boundary that separates the region where complete synchronization occurs on the space of parameters $(p,\epsilon)$ for the autonomous system Eq.~(\ref{auto_syst}), with $H$ given by Eq.~(\ref{partial}) for any value of $q$, coincides with the stability boundary $\Lambda_x=0$ in Fig.~\ref{fig2} for the driven system with $g=f$.

\begin{figure}[h]
\centerline{
\includegraphics[scale=0.7]{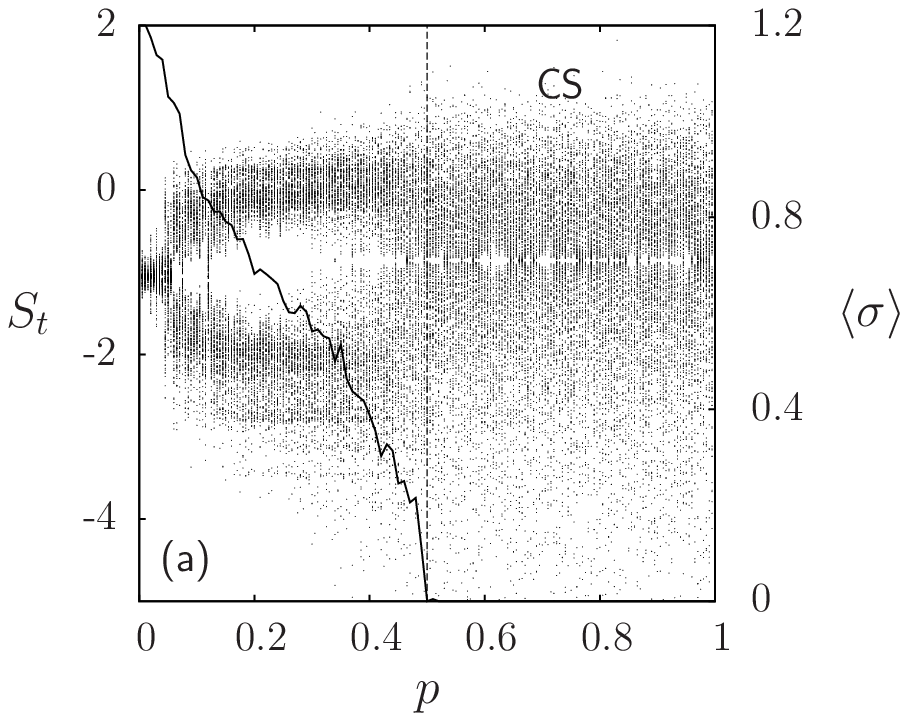}}
\vspace{1mm}
\centerline{
\includegraphics[scale=0.7]{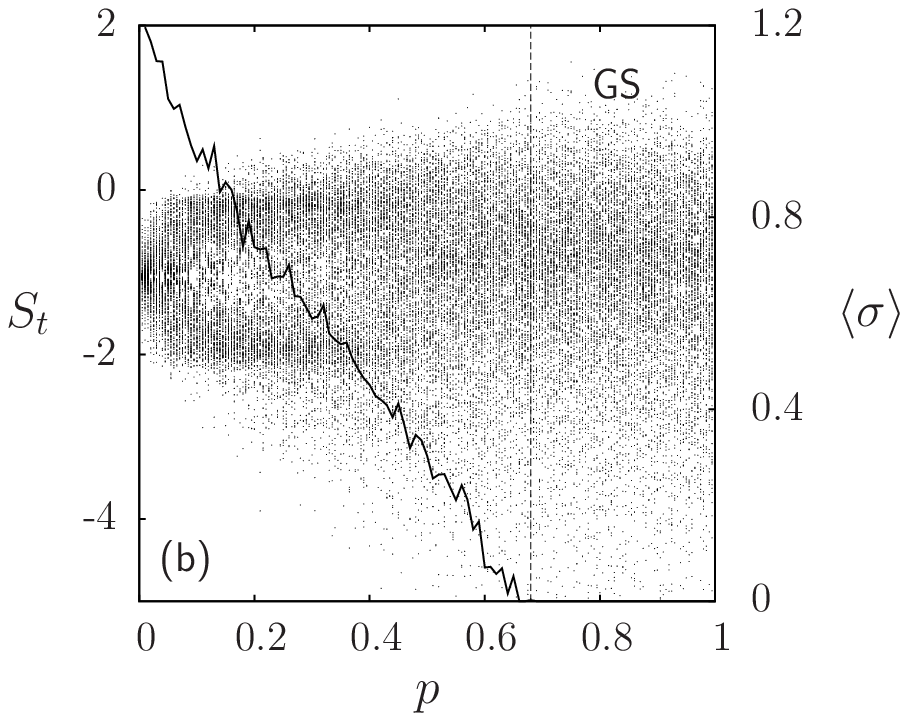}}
\caption{Left vertical axis: bifurcation diagram of $S_t$ as a function of
$p$ for the homogeneously coupled autonomous system Eq.~(\ref{auto_syst}).  Right vertical axis: $\langle \sigma \rangle$ vs $p$, continuous line. For each value of $p$, $S_t$ and $\langle\sigma\rangle$ were calculated at each time step during a run starting from random initial conditions on the local maps, uniformly distributed on the interval $[-10,3]$, after discarding the transients. (a)  $H_t$ given by  Eq.~(\ref{partial}) with $q=0.5N$; the region where the complete synchronization state occurs is labeled CS. (b) $H_t$ given by Eq.~(\ref{shift}) with $k=1.2$ and $q=0.5N$; the region where generalized synchronization occurs is indicated by GS.  In both cases, $f(x)=-0.7+\log|x|$, $\epsilon=0.5$, and $N=10^4$.}
\label{fig4}
\end{figure}

Figure~\ref{fig4}(a) shows the bifurcation diagram of the mean field $S_t$
as well as $\langle \sigma \rangle$ as functions of the probability $p$ for the autonomous system with the coupling function $H$ given by Eq~(\ref{partial}), with a fixed value of $\epsilon$. The mean field $S_t$ is chaotic for all values of $p$. There is a critical value of $p$ above which $\langle \sigma \rangle =0$ and $f(x_t^i)=f(x_t)=H$, indicating that the system is completely synchronized in a chaotic state for that range of the probability of interactions.

For other functional forms of the coupling function $H$ it is possible to find generalized synchronized states in the autonomous system Eq.~(\ref{auto_syst}).
For example, consider the coupling function
\begin{equation}
\label{shift}
 H( x^j_t:j \in Q_t)= k + \frac{1}{q} \sum^q_{j=1} f(x^j_t),
\end{equation}
where $k$ is a constant and $q<N$ elements are chosen at random at each time $t$.

Figure~\ref{fig4}(b) shows both the bifurcation diagram of $S_t$
and the quantity $\langle \sigma \rangle$ as functions of the probability $p$ for the autonomous system  Eq.~(\ref{auto_syst}) with the coupling function $H$ given by Eq~(\ref{shift}), with a fixed value of $\epsilon$. Again, there is a value of $p$ above which $\langle \sigma \rangle$ vanishes, signaling the onset of a synchronized chaotic state. In this region $f(x_t^i)=f(x_t) \neq H=k+f(x_t)$, independently of the value of $q$,  corresponding to 
generalized synchronization in the autonomous system Eq.~(\ref{auto_syst}).

We have verified that the regions of the parameter space $(p,\epsilon)$ where complete or generalized synchronized states emerge in the autonomous system with intermittent global interactions, Eq.~(\ref{auto_syst}), are the same regions that correspond to those states for the single driven map in Figure~\ref{fig1}.  

\section{Autonomous systems with heterogeneous coupling}
In analogy to a system of heterogeneously driven maps, Eq.~(\ref{part_drive}), we can consider
the following autonomous coupled map system,
\begin{equation}
x^i_{t+1} = \left\lbrace 
\begin{array}{ll}
(1-\epsilon)f(x^i_t)+ \epsilon H( x^j_t:j \in Q_t), & \text{prob. $p$},\\
f(x_t^i), & \text{prob. $(1-p)$}.
\end{array}
\right. 
\label{pauto_syst}
\end{equation}
The parameter $p$ is the probability of interaction of an element with the coupling function at time $t$. Thus the average
fraction of fully connected elements at any given time is $p$. For $p=1$, the autonomous systems Eq.~(\ref{pauto_syst}) and Eq.~(\ref{auto_syst}) are identical. 

Not all the maps in the system Eq.~(\ref{pauto_syst}) experience the same global interaction at a given time. 
However, at the local level, each map in this autonomous system is subject, on the average, to the same 
coupling function $H$ with probability $p$ over long times. Similarly, each map in the autonomous homogeneously coupled system Eq.~(\ref{auto_syst}) experiences the coupling $H$ with a probability $p$ in time.  
This is also analogous to the condition of the local dynamics in the heterogeneously driven system of maps, Eq.~(\ref{part_drive}), if $g$ exhibits the same temporal evolution as $H$. 
Then, the synchronization behavior of the system Eq.~(\ref{pauto_syst}) should also be similar to the
behavior of the single map driven, Eq.~(\ref{2dmap}) subject to a drive that satisfies $g(y_t)=H$ for asymptotic times.  

When the autonomous system Eq.~(\ref{pauto_syst}) reaches a synchronized state, we have $f(x_t^i)=f(x_t)$. However, this synchronized solution exists only if $H=f(x_t)$. Therefore, as in the case of the heterogeneously driven system, only complete synchronization $(x_t^i)=f(x_t)=H$ arises in the system Eq.~(\ref{pauto_syst}).

\begin{figure}[h]
\centerline{
\includegraphics[scale=0.7]{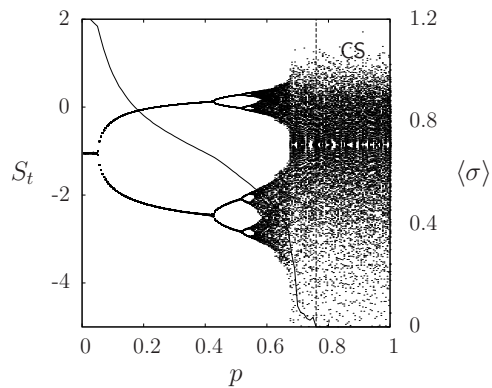}}
\caption{Left vertical axis: bifurcation diagram of $S_t$ as a function of
$p$ for the partially coupled autonomous system Eq.~(\ref{pauto_syst}) with $H_t$ defined in Eq.~(\ref{partial}) and $q=0.5N$. The region where the complete synchronization state occur is labeled CS. Right vertical axis: $\langle \sigma \rangle$ vs $p$, continuous line. For each value of $p$, $S_t$ and $\langle\sigma\rangle$ were calculated at each time step during a run starting from random initial conditions on the local maps, uniformly distributed on the interval $[-10,3]$, after discarding the transients. Here, $f(x)=-0.7+\log|x|$, $\epsilon=0.4$,  and $N=10^5$.}
\label{fig5}
\end{figure}

As an example of complete synchronization in the autonomous system Eq.~(\ref{pauto_syst}), we choose the form Eq.~(\ref{partial}) as the coupling function $ H( x^j_t:j \in Q_t)$  
Figure~\ref{fig5} shows $S_t$ and the quantity $\langle \sigma \rangle$ as functions of the probability $p$, for a fixed value of $\epsilon$. For the given values of parameters, there is a threshold value of $p$ at which $\langle \sigma \rangle= 0$, indicating that the elements become synchronized; in this case, complete synchronization, $f(x_t^i)=f(x_t) = H$, independently of the value of $q$. Thus, the reinjection of a coupling function containing partial information of the system to a fraction of randomly selected elements suffices to achieve complete synchronization in the system. 

Beyond the common regions for complete or generalized synchronization, the driven and the autonomous systems may exhibit different collective behaviors in the parameter space.
This is particularly manifested in the autonomous system with inhomogeneous coupling. 
Besides complete synchronization, the instantaneous mean field $S_t$ in Figure~\ref{fig5} (left vertical axis) shows more complex collective behaviors arising in the system Eq.~(\ref{pauto_syst}). For the parameter value $b=-0.7$, the local maps are
chaotic. However, for some range of the parameters $p$ and $\epsilon$, $S_t$ reveals the existence of
global periodic attractors. In Figure~\ref{fig5}, collective periodic states of $S_t$ at a given
value of the parameter $p$ appear as sets of small vertical segments which correspond to intrinsic
fluctuations of the periodic orbits of the mean field. Global attractors of period $1$, $2$, $4$, and $8$ are observed in the
mean field as the parameter $p$ is varied, and they become better defined as the system size $N$ is increased. The amplitudes
of the global periodic motions of $S_t$ do not decrease with an increase of $N$. As a consequence, the variance of $S_t$ itself does not decay as $N^{−1}$ with increasing $N$ but rather it saturates at some constant value related to the amplitude of the collective period. This is a phenomenon of nontrivial collective behavior, where macroscopic quantities in a spatiotemporal
dynamical system exhibit spontaneous ordered evolution coexisting with local chaos (Chat\'e \& Manneville 1998). 
This phenomenon arises in the heterogeneously coupled autonomous system, rather than in the homogeneously coupled one. 
In the first case, the coupling function presents spatial fluctuations than can be interpreted as a distributed noise. In fact, 
it has been shown that local noise can enhance the emergence of ordered collective behavior in coupled chaotic map systems (Shibata {\it et al.} 1999). Note that the emergence of collective periodic behavior in our case shown in Figure~\ref{fig5} cannot be attributed to the presence of periodic windows in the local dynamics since the logarithmic map possesses robust chaos for $b \in [−1, 1]$ (Alvarez-Llamoza {et al.} 2007). 

\begin{figure}[h]
\centerline{
\includegraphics[scale=0.54]{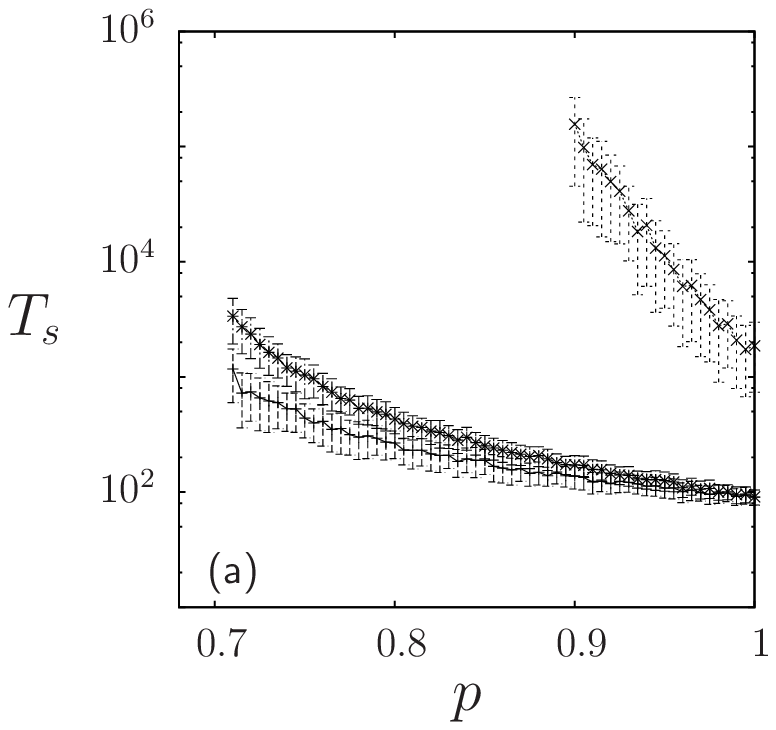}
\hspace{1mm}
\includegraphics[scale=0.54]{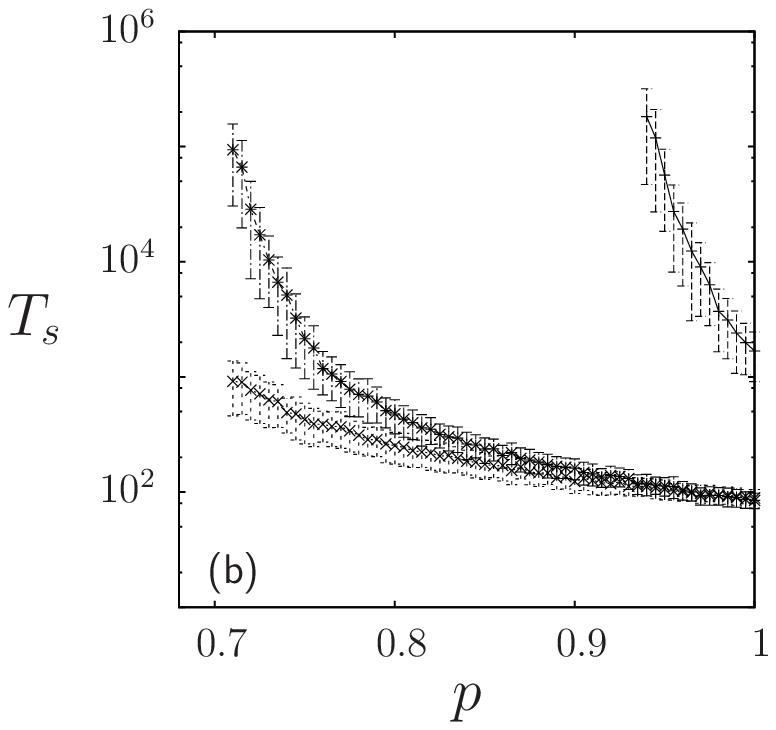}}
\caption{Average time $T_s$ to reach synchronization as a function of the probability $p$. The error bars on each curve correspond to standard deviations resulting from $100$ realizations of random initial conditions for each value of $p$. (a) Driven systems with fixed $\epsilon=0.4$ and $N=10^4$.
Top curve: homogeneously driven system, Eq.~(\ref{int_drive}), with $f=-0.7+\log|x|$, $g=0.5+\log|y|$ (generalized synchronization).
Middle curve: heterogeneously driven system, Eq.~(\ref{part_drive}), with $f=g=-0.7+\log|x|$ (complete synchronization).
Bottom curve:  homogeneously driven system, Eq.~(\ref{int_drive}),  with $f=g=-0.7+\log|x|$ (complete synchronization).
(b) Autonomous systems with fixed $\epsilon=0.4$ and $N=10^4$. 
Top curve: autonomous homogeneously coupled system, Eq.~(\ref{auto_syst}), with $H$ given by Eq~(\ref{shift}), $q=0.5N$
and $k=1.2$ (generalized synchronization). Middle curve: autonomous heterogeneously coupled system, Eq.~(\ref{pauto_syst}), 
with $H$ given by Eq.~(\ref{partial}), $q=0.5N$ (complete synchronization). Bottom curve: autonomous homogeneously coupled system, Eq.~(\ref{auto_syst}), with $H$ given by Eq.~(\ref{partial}), $q=0.5N$ (complete synchronization).}
\label{fig6}
\end{figure}

Although complete or generalized synchronization of chaos in both, driven and autonomous systems, can be characterized from the knowledge of the dynamical response of a single driven map, the transient behavior to reach such a state may be different in each case. Figure~(\ref{fig6}) shows the average time $T_s$ required to attain complete or generalized synchronization as a function of $p$ in the driven and the autonomous systems considered in this paper. 
The time $T_s$ for generalized synchronization is larger than that for complete synchronization in an extended system subject to a field, either external or endogenous. In general, a homogeneous, synchronous field acting on a system is more efficient for achieving complete synchronization. 
The curves $T_s$ vs. $p$ for complete synchronization in either a driven or an autonomous system subject to a homogeneous field are practically indistinguishable. The same curve is also obtained for a single driven map possesing similar parameter values, as one may expect from the local dynamics analogy. Varying the size of the system $N$ does not significantly affect the curves in Figure~(\ref{fig6}).

Note that for achieving synchronization in either the homogeneous or heterogeneous autonomous system, the coupling function $H$ does not need to depend on all the internal variables; nor must it be active for all times or
act simultaneously on all the elements.   
What becomes essential for synchronization is that all the elements in an autonomous system share some minimal information, on the average, over time. This is the same process that leads to synchronization in the associated driven systems Eq.~(\ref{int_drive}) and  Eq.~(\ref{part_drive}), respectively, and it is also the mechanism for synchronization in the single driven map, Eq.~(\ref{2dmap}). Thus, the nature of the common
influence, either external or internal, affecting the local units becomes irrelevant for the emergence of synchronization in driven or autonomous systems. 

\section{Conclusions}
We have shown that the synchronization behavior of system of chaotic maps subject to either an external forcing or a coupling function of their internal variables can be inferred from the behavior of a single  element in the system.  
The local dynamics can be seen as a single driven map. When this single drive-response system reaches either complete or generalized synchronization, the local analogy implies that an ensemble of identical maps subject to a common field, either external or internal, that behaves as this drive, should also synchronize in a similar manner. 

By considering the conditions for stable synchronization in a single driven-map model with minimal ingredients, we have found some minimal, equivalent conditions for the emergence of complete and generalized chaos synchronization in both driven and autonomous associated systems. 

For the homogeneously driven or coupled autonomous systems, the sharing of the influence by the elements takes place only a fraction of the time. In the case of the heterogeneously driven or coupled autonomous system, 
the influence is shared only by a fraction of elements at any time.
Our results show that the presence of a common drive or a coupling function for all times is not indispensable for reaching synchronization in an extended system of chaotic oscillators, nor is the simultaneous sharing of the drive or the coupling function by all the elements in the system. In the case of an autonomous system, the coupling function does not need to depend on all the internal variables for achieving synchronization and, in particular, its functional form is not crucial for generalized synchronization.  What matters for reaching synchronization in a system is
the sharing of some minimal information by the elements, on the average, over long times, independently of its source.
This result suggests that these systems possess an ergodicity property.

Future extensions of this work include the investigation of some quantity, such as the transfer entropy (Schreiber, 2000) for measuring this minimal amount of information required for either type of synchronization, and
the possibility of prediction of other forms of collective behaviors or patterns observed in dynamical networks, from the knowledge of the occurrence of these phenomena in an associated single driven-response system. 

\section*{References}
\begin{description}
 
\item  Abarbanel, H. D. I., Rulkov, N F \&  Sushchik, M. M. 1996 Generalized synchronization of chaos: The auxiliary system approach. {\it Phys. Rev. E} {\bf 53}, 4528-4535. 
\item  Alvarez-Llamoza, O. \&  Cosenza, M. G. 2008 Generalized synchronization of chaos in autonomous systems. {\it Phys. Rev. E} {\bf 78}, 046216-1-046216-6. 
\item  Alvarez-Llamoza, O., Tucci, K., Cosenza, M. G., \& Pineda, M. 2007
Random global coupling induces synchronization
and nontrivial collective behavior in networks
of chaotic maps. {\it Eur. Phys. J. Special Topics} {\bf 143}, 245-247. 
\item  Argyris, A., Syvridis, D.,  Larger, L., Annovazzi-Lodi, V.,  Colet, P., Fischer, I., 
Garcia-Ojalvo, J., Mirasso, C. R., Pesquera, L. \& Shore, K. A. 2005  Chaos-based communications at high bit rates using commercial fibre-optic links. {\it Nature (London)} {\bf 438}, 343-346. 
\item  Boccaletti, S.,  Kurths, J., Osipov, G., Valladares, D. L., \&  Zhou, C. S. 2002 The synchronization of chaotic systems. {\it Phys. Rep.} {\bf 366}, 1-101. 
\item  Chat\'e, H. \& Manneville, P., 1992  Emergence of effective low-dimensional dynamics in the macroscopic behaviour of coupled map lattices. {\it Europhys. Lett.} {\bf 17}, 291-296. 
\item  De Monte, S., d'Ovidio, F., Dan\o{}, S., S\o{}rensen, P. G. 2007 Dynamical quorum sensing: population density encoded in cellular dynamics. {\it Proc. Natl. Acad. Sci.} {\bf 104}, 18377-18381.  
\item  Hunt, B. R.,  Ott, E. \& Yorke, J. A.  1997 Differentiable generalized synchronization of chaos. {\it Phys. Rev. E} {\bf 55}, 4029-4034. 
\item  Kaneko, K. 1989 Chaotic but regular posi-nega switch among coded attractors by cluster-size variation {\it Phys. Rev. Lett.} {\bf 63}, 219-223.  
\item  Kapitaniak, T.,   Wojewoda, J.\& Brindley, J. 1996 Synchronization and desynchronization in quasi-hyperbolic chaotic systems. {\it Phys. Lett. A} {\bf 210}, 283-289.  
\item  Kawabe, T. \& Kondo Y. 1991 Fractal transformation of the one-dimensional chaos produced by logarithmic map. {\it Prog. Theor. Phys.} {\bf 85}, 759-759.  
\item  Manrubia, S. C.,  Mikhailov, A. S. \& Zanette,  D. H. 2004 \textit{Emergence of dynamical order: synchronization phenomena in complex systems}. Singapore: World Scientific.  
\item  Masoller, C.  \&  Marti, A. 2005 Random delays and the synchronization of chaotic maps. {\it Phys. Rev. Lett.} {\bf 94}, 134102-1-134102-4.
\item  Newman, M. E. J.,  Barabasi, A. L. \&  Watts, D. J. 2006 {\it The Structure and dynamics of networks}. Princeton, NJ:
Princeton University Press. 
\item  Parlitz, U. \&  Kokarev, L. 1999. In {\it Handbook of Chaos Control} (ed.  H. Schuster), p. 271.  Weinheim: Wiley-VCH. 
\item  Parravano, A. \&  Cosenza, M. G. 1998  Driven maps and the emergence of ordered collective behavior in globally coupled maps. {\it Phys. Rev. E} {\bf 58}, 1665-1671. 
\item  Pecora, L. M. \& Carroll, T. L. 1990 Synchronization in chaotic systems. {\it Phys. Rev. Lett.} {\bf 64}, 821-824. 
\item  Pikovsky, A., Rosenblum, M.  \&  Kurths, J. 2002 \textit{Synchronization: a
universal concept in nonlinear sciences}.  Cambridge, UK: Cambridge University Press.  
\item  Rulkov, N. F., Sushchik, M. M., Tsimring, L. S. \& Abarbanel,  H. D. I. 1995 Generalized synchronization of chaos in directionally coupled systems. {\it Phys. Rev. E} {\bf 51}, 980-994. 
\item  Schreiber, T. 2000 Measuring information transfer. {\it Phys. Rev. Lett.} {\bf 85}, 461-464. 
\item  Shibata, T., Chawanya, T., \& Kaneko, K. 1999 Noiseless collective motion out of noisy chaos. {\it Phys. Rev. Lett.} {\bf 82}, 4424-4427. 
\item  Taylor, A.F., Tinsley, M. R.,  Wang, F., Huang, Z. \& Showalter, K. 2009 Dynamical quorum sensing and synchronization in large populations of chemical oscillators. {\it Science} {\bf 323}, 614-617.  
\item  Uchida, A., Rogister, F., Garcia-Ojalvo, J. \& Roy,  R. 2005  Synchronization and communication with chaotic laser systems. {\it Prog. Opt.} {\bf 48}, 203-341. 
\item  Waller, I. \& Kapral, R. 1984 Spatial and temporal structure in systems of coupled nonlinear oscillators. {\it Phys. Rev. A} {\bf 30}, 2047-2055.  
\item  Wang, W., Kiss, I. Z.,  \& Hudson, J. L. 
 2000 Experiments on arrays of globally coupled chaotic electrochemical oscillators: synchronization and clustering.
{\it Chaos} {\bf 10}, 248-256. 
\item  Zhou, B. B.  \& Roy, R. 2007 Isochronal synchrony and bidirectional communication with delay-coupled nonlinear oscillators. {\it Phys. Rev. E} \textbf{75},  026205-1- 026205-5. 

\end{description}
\end{document}